\documentclass[a4paper,twocolumn,10pt,accepted=2026-11-01]{quantumarticle}
\pdfoutput=1

\usepackage{geometry}
\usepackage[english]{babel}
\usepackage[T1]{fontenc}
\usepackage{amsmath}
\usepackage{hyperref}
\usepackage{amsfonts}
\usepackage{bbm}
\usepackage{caption}
\usepackage{mathtools}
\usepackage{xspace}
\usepackage[numbers,sort&compress]{natbib}

\begin{document}

\title{Lower-depth local unitary encoding circuits for the surface code}
\author{Jahan Claes}
\email{jahan@logiqal.com}
\affiliation{Logiqal Inc., Monmouth Junction, NJ 08852, USA}
\begin{abstract}
The surface code is the most studied error-correcting code thanks to its high threshold, simple decoding, and locality in two dimensions (2D). A key component of any code is its encoding circuit, which maps an unencoded state to the corresponding encoded state. The best previous surface code encoding circuit compatible with 2D local connectivity requires depth $2d$ to encode distance-$d$ surface codes. This paper presents depth $d$ encoding circuits for the rotated surface code. Our circuit is constructed inductively from circuits that grow the code from $d$ to $(d+2)$. We prove that depth $d+O(1)$ is optimal for inductively constructed circuits.
\end{abstract}
\maketitle

\section{Introduction}

The surface code is a promising candidate for fault-tolerant quantum computation in 2D local architectures~\cite{bravyi1998quantum,dennis2002topological,fowler2012surface,kitaev1997quantum} and is also one of the simplest models of topological order~\cite{kitaev2003fault}. For any code, it is useful to be able to encode arbitrary states $|\psi\rangle$ using a unitary circuit. On the quantum computing side, encoding circuits are useful for magic state injection~\cite{rodriguezExperimentalDemonstrationLogical2024,gidneyMagicStateCultivation2024,claes2025cultivation,puri2025cultivation,aws2025cultivation} and early demonstrations of error-detected computation~\cite{bluvsteinLogicalQuantumProcessor2024,rodriguez2412experimental}. On the quantum simulation side, encoding circuits have been used to realize topological order without mid-circuit measurements and feed-forward~\cite{satzingerRealizingTopologicallyOrdered2021}.

It has long been known that any 2D local encoding circuit for the surface code must have depth proportional to the distance $d$ of the surface code~\cite{bravyi2006lieb} (encoding circuits that achieve \textit{logarithmic} depth have been proposed~\cite{vidalClassQuantumManyBody2008,aguadoEntanglementRenormalizationTopological2008,liao2021graph,tsai2025unitaryencodersurfacecodes}, but these necessarily use nonlocal gates). The current best 2D local encoding circuits for the surface code are depth-$2d$, and use nearest-neighbor and diagonal connectivity on a square grid of qubits~\cite{higgott2021optimal,chenQuantumCircuitsToric2024}.

In this note, we present depth-$d$ encoding circuits on a square grid with only nearest-neighbor connectivity for the rotated surface code, halving the circuit depth and reducing the connectivity requirements compared to previous work. Our circuit uses $6d+O(1)$ two-qubit gates, compared to $8d+O(1)$ for the previous best local circuit~\cite{higgott2021optimal}. Like~\cite{higgott2021optimal}, our circuit is defined inductively: we give circuits for growing a surface code from $d$ to $(d+2)$ for arbitrary $d$. We show that any circuit similarly defined inductively will always have depth at least $d$, so our circuits are in some sense optimal. However, we leave open the possibility that a circuit not built out of a series of $d\rightarrow (d+1)$ and/or $d\rightarrow(d+2)$ circuits could achieve lower depth.

\section{Our circuits}

\begin{figure}
\includegraphics[width=\columnwidth]{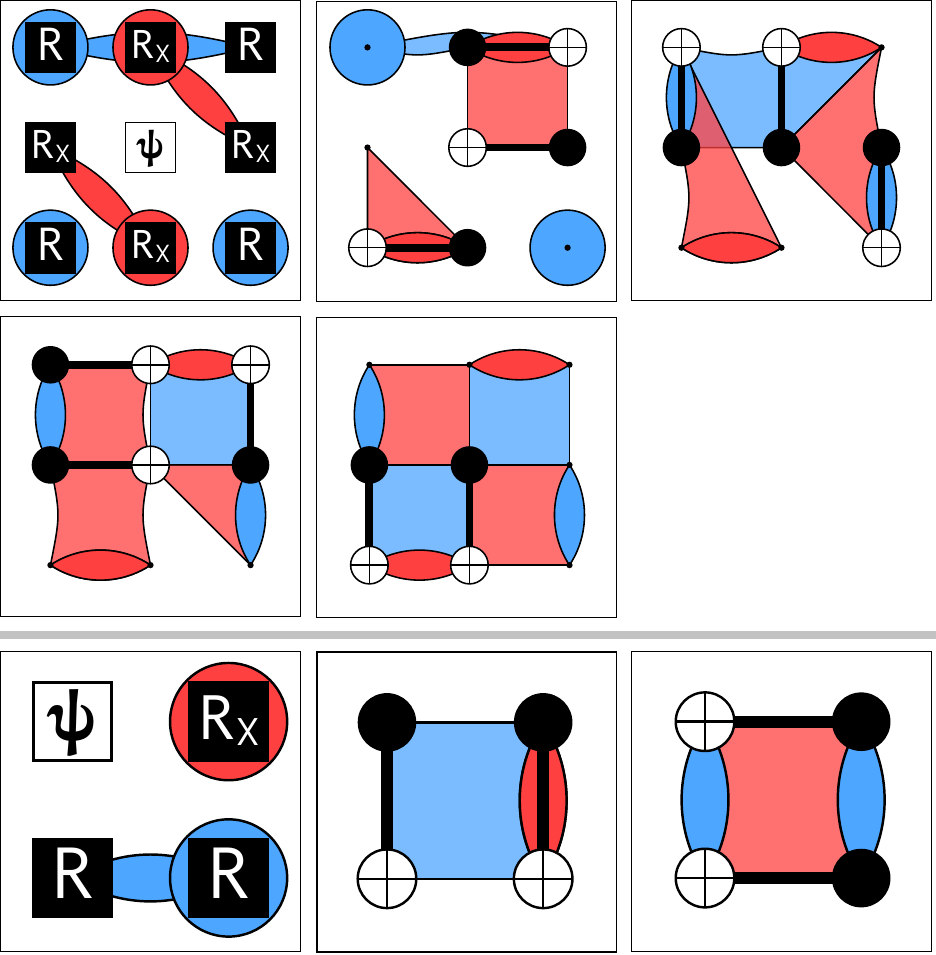}
\caption{Encoding an initial state $|\psi\rangle$ into small $d=2,3$ surface codes as the starting step of our inductive construction. Top: A depth-$4$ encoding circuit for $d=3$ (\href{https://algassert.com/crumble\#circuit=Q(0,0)0;Q(0,1)1;Q(0,2)2;Q(1,0)3;Q(1,1)4;Q(1,2)5;Q(2,0)6;Q(2,1)7;Q(2,2)8;R_6_0_2_8;RX_3_1_5_7;S_DAG_4;MARKX(0)3_7;MARKX(1)3;MARKX(2)1_5;MARKX(3)5;MARKZ(4)6_0;MARKZ(5)0;MARKZ(6)2;MARKZ(7)8;TICK;CX_7_4_3_6_5_2;TICK;CX_4_3_7_8_1_0;TICK;CX_1_4_0_3_7_6;TICK;CX_4_5_1_2}{open in Crumble}). Ref~\cite{tsai2025unitaryencodersurfacecodes} provides an alternative depth-$4$ encoding circuit for $d=3$. Bottom: A depth-$2$ encoding circuit for $d=2$ (\href{https://algassert.com/crumble\#circuit=Q(0,0)0;Q(0,1)1;Q(1,0)2;Q(1,1)3;R_1_3;RX_2;MARKX(2)2;MARKZ(1)1_3;MARKZ(3)3;TICK;CX_2_3_0_1;TICK;CX_2_0_3_1}{open in Crumble}).}
\label{fig:Encoding}
\end{figure}

\begin{figure*}[!t]
    \includegraphics[width=\textwidth]{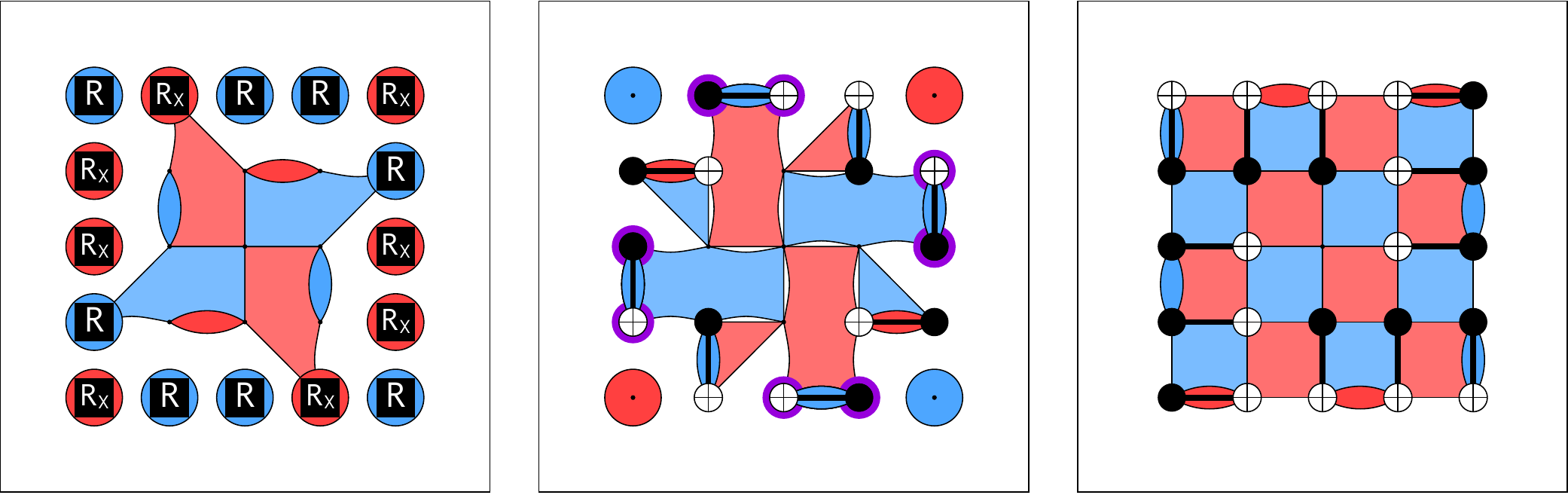}
    \ 
    
    \includegraphics[width=\textwidth]{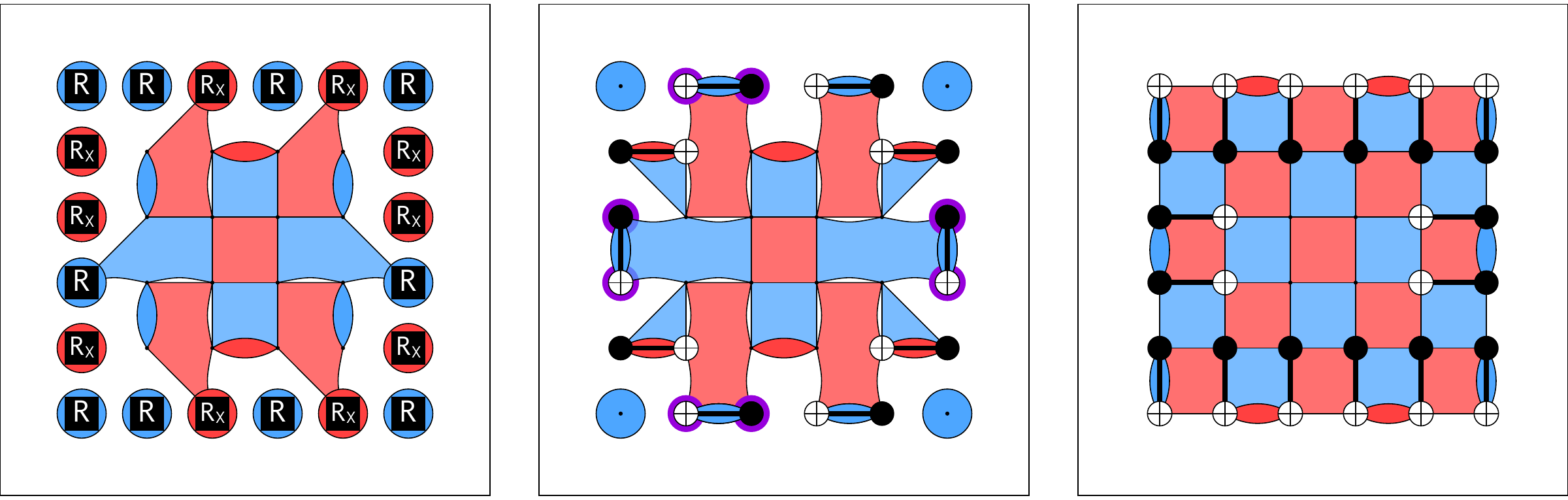}
    \caption{Top: The circuit to grow from $d=3$ to $d=5$. All $d\rightarrow (d+2)$ circuits with $d$ odd are a version of this circuit with the purple qubits repeated. \href{https://algassert.com/crumble\#circuit=Q(0,0)0;Q(0,1)1;Q(0,2)2;Q(0,3)3;Q(0,4)4;Q(0,5)5;Q(0,6)6;Q(1,0)7;Q(1,1)8;Q(1,2)9;Q(1,3)10;Q(1,4)11;Q(1,5)12;Q(1,6)13;Q(2,0)14;Q(2,1)15;Q(2,2)16;Q(2,3)17;Q(2,4)18;Q(2,5)19;Q(2,6)20;Q(3,0)21;Q(3,1)22;Q(3,2)23;Q(3,3)24;Q(3,4)25;Q(3,5)26;Q(3,6)27;Q(4,0)28;Q(4,1)29;Q(4,2)30;Q(4,3)31;Q(4,4)32;Q(4,5)33;Q(4,6)34;Q(5,0)35;Q(5,1)36;Q(5,2)37;Q(5,3)38;Q(5,4)39;Q(5,5)40;Q(5,6)41;Q(6,0)42;Q(6,1)43;Q(6,2)44;Q(6,3)45;Q(6,4)46;Q(6,5)47;Q(6,6)48;MARKX(0)16_23_24_17;MARKX(1)24_25_32_31;MARKX(2)25_18;MARKX(3)23_30;MARKZ(4)16_17;MARKZ(5)17_24_25_18;MARKZ(6)24_31_30_23;MARKZ(7)31_32;TICK;R_37_40_26_8_22_11_19_29;RX_9_10_38_39_33_15_12_36;MARKX(0)15;MARKX(1)33;MARKX(12)9;MARKX(13)10;MARKX(14)12;MARKX(15)36;MARKX(16)38;MARKX(17)39;MARKX(8)15;MARKX(9)33;MARKZ(10)11;MARKZ(11)37;MARKZ(18)8;MARKZ(19)19;MARKZ(20)22;MARKZ(21)26;MARKZ(22)29;MARKZ(23)40;MARKZ(5)11;MARKZ(6)37;TICK;CX_39_32_38_37_33_26_9_16_10_11_15_22_18_19_30_29;TICK;CX_25_26_32_33_39_40_38_31_37_30_10_17_11_18_9_8_16_15_23_22_12_19_36_29;TICK;TICK;R_0_14_28_35_43_45_48_34_20_13_5_3;RX_7_1_2_4_6_27_41_47_46_44_42_21;MARKX(12)7;MARKX(17)41;MARKX(2)27;MARKX(24)7;MARKX(25)21;MARKX(26)42;MARKX(27)44;MARKX(3)21;MARKX(35)27;MARKX(41)6;MARKX(42)41;MARKX(43)47;MARKX(44)46;MARKX(46)1;MARKX(47)2;MARKX(48)4;MARKZ(19)5;MARKZ(22)43;MARKZ(28)0;MARKZ(29)14;MARKZ(30)28;MARKZ(31)35;MARKZ(32)43;MARKZ(33)45;MARKZ(34)34;MARKZ(36)20;MARKZ(37)13;MARKZ(39)5;MARKZ(4)3;MARKZ(40)48;MARKZ(45)3;MARKZ(7)45;TICK;CX_1_8_7_14_21_28_36_35_44_43_46_45_47_40_41_34_27_20_12_13_4_5_2_3;TICK;CX_1_0_8_7_15_14_22_21_29_28_47_48_40_41_33_34_26_27_19_20_42_35_43_36_44_37_45_38_46_39_2_9_3_10_4_11_5_12_6_13}{Open odd-$d$ circuit in Crumble}. 
    Bottom: The circuit to grow from $d=4$ to $d=6$.  All circuits with $d$ even are a version of this circuit with the purple qubits repeated (or for $d=2\rightarrow 4$, removed). \href{https://algassert.com/crumble\#circuit=Q(0,0)0;Q(0,1)1;Q(0,2)2;Q(0,3)3;Q(0,4)4;Q(0,5)5;Q(1,0)6;Q(1,1)7;Q(1,2)8;Q(1,3)9;Q(1,4)10;Q(1,5)11;Q(2,0)12;Q(2,1)13;Q(2,2)14;Q(2,3)15;Q(2,4)16;Q(2,5)17;Q(3,0)18;Q(3,1)19;Q(3,2)20;Q(3,3)21;Q(3,4)22;Q(3,5)23;Q(4,0)24;Q(4,1)25;Q(4,2)26;Q(4,3)27;Q(4,4)28;Q(4,5)29;Q(5,0)30;Q(5,1)31;Q(5,2)32;Q(5,3)33;Q(5,4)34;Q(5,5)35;MARKX(4)14_20_21_15;MARKZ(3)15_14;MARKZ(5)21_20;TICK;R_7_13_25_10_16_28;RX_19_26_27_22_8_9;MARKX(0)8;MARKX(2)26;MARKX(33)19;MARKX(34)22;MARKX(4)19_22;MARKX(6)9;MARKX(8)27;MARKZ(1)13;MARKZ(29)7;MARKZ(30)10;MARKZ(31)25;MARKZ(32)28;MARKZ(7)16;TICK;CX_8_14_19_13_26_20_9_15_27_21_22_16;TICK;CX_8_7_14_13_20_19_26_25_9_10_15_16_21_22_27_28;TICK;TICK;R_3_33_0_5_35_30_6_11_23_18;RX_2_32_1_4_31_34_12_24_29_17;MARKX(0)12;MARKX(19)2;MARKX(2)24;MARKX(20)32;MARKX(21)1;MARKX(22)4;MARKX(23)31;MARKX(24)34;MARKX(25)12;MARKX(26)24;MARKX(27)29;MARKX(28)17;MARKX(6)17;MARKX(8)29;MARKZ(10)33;MARKZ(11)0;MARKZ(12)5;MARKZ(13)35;MARKZ(14)30;MARKZ(15)6;MARKZ(16)11;MARKZ(17)23;MARKZ(18)18;MARKZ(3)3;MARKZ(5)33;MARKZ(9)3;TICK;CX_1_7_34_28_31_25_4_10_2_3_32_33_12_6_24_18_29_23_17_11;TICK;CX_32_26_33_27_3_9_2_8_7_6_1_0_13_12_19_18_25_24_31_30_4_5_10_11_16_17_22_23_28_29_34_35}{Open even-$d$ circuit in Crumble}.}
    \label{fig:Circuits}
\end{figure*}

We illustrate our circuits using Stim visualizations, including the instantaneous $X$ ($Z$) stabilizers shown in red (blue). There is always an arbitrary basis choice when defining an instantaneous stabilizer group, as we can always replace stabilizers $\{S_i,S_j\}$ with $\{S_i,S_iS_j\}$. For our illustrations, we choose the basis to be the usual surface code stabilizers in the \textit{last} step. As a result, our stabilizers in the \textit{first} step appear somewhat arbitrary.

Our circuits will grow a surface code from $d$ to $(d+2)$ in depth $2$. To start the inductive construction for odd (even) $d$, we encode into a $d=3$ ($d=2$) surface code using a circuit of depth $4$ (depth $2$). Thus, our overall circuit for distance $d$ will have depth $d+[d\text{ mod }2]$. The encoding circuits are illustrated in Fig.~\ref{fig:Encoding}.

We now present our growth circuits. We use two distinct circuit families, one for $d$ odd and one for $d$ even. We illustrate the circuits for $d=3\rightarrow 5$ and $d=4\rightarrow 6$ in Fig.~\ref{fig:Circuits}. For $d=2$, we eliminate the two purple-highlighted qubits on each of the four sides. For $d>4$, we repeat the two purple-highlighted qubits as many times as necessary to reach the desired $d$. Our $d\rightarrow(d+2)$ circuit uses $(6d+6)$ two-qubit gates for $d$ odd, compared to $(8d+4)$ in~\cite{higgott2021optimal}. For $d$ even our circuit uses $(6d+5)$ gates, which was not considered in~\cite{higgott2021optimal}.

While our circuits are straightforward to understand pictorally, they are cumbersome to describe algorithmically. We include a program to generate Stim circuits for arbitrary-distance versions of our codes, and to verify that our circuits indeed prepare the surface code, at~\cite{claes_2026_19502022}.

Our circuit construction was enabled by the interactive stabilizer circuit editor Crumble. Previous work on unitary circuits likely did not discover our low-depth construction because Crumble was not available at the time.

\section{Optimality of depth-$d$ inductive circuits}

We now show that any Clifford circuit that grows a surface code from $d$ to $(d+2)$ must have depth at least two. It immediately follows that any injection circuit constructed from a series of $d\rightarrow (d+1)$ and/or $d\rightarrow (d+2)$ circuits necessarily has depth $d+O(1)$.

To grow a surface code from $d$ to $(d+2)$, we prepare $4(d+1)$ fresh qubits in product states and then perform a unitary circuit. Before the unitary circuit, there are $4(d+1)$ independent single-qubit stabilizers. If the unitary circuit is depth-one, each single-qubit stabilizer evolves into a single-qubit or two-qubit stabilizer. However, every basis of stabilizers of a $(d+2)$ surface code has at most $2(d+1)$ two-qubit stabilizers and no single-qubit stabilizers. Therefore, a unitary circuit sending $d\rightarrow (d+2)$ cannot be depth-one.

\section{Conclusion}

One of the most basic ingredients of an error-correcting code is its injection circuit. Our surface code injection circuit is half the depth of the best previous 2D local circuit, is realizable on a square grid, and is the lowest-depth circuit possible for growing a surface code from $d$ to $(d+2)$. Our circuits could likely be used for larger realizations of topological order than previous work~\cite{satzingerRealizingTopologicallyOrdered2021}. Small versions of our circuits have also already been used in magic state cultivation for the surface code~\cite{claes2025cultivation}. We leave the question of further applications open to future work.

\section{Author contribution statement}

This manuscript represents work done entirely by the author, with no assistance or contribution from from any LLM. However, in the second version of this manuscript, I have included a link to a Python function that generates injection circuits in Stim for arbitrary distance $d$~\cite{claes_2026_19502022}; this Python file was generated by Claude based on a set of example circuits. 

As an aside, when I first posted this paper to the arXiv in 2025, it was still the case that clever Clifford circuit constructions were a valuable (and fun!) part of QEC research. Revisiting this paper after acceptance in Quantum in late 2026, it is clear that humans are no longer competitive with LLMs in this area. Designing these circuits requires lots of iteration but they are easy to validate. I recently had reason to want a similar injection circuit for the color code; what would have been a week of work and a cute follow-up paper a year ago is now a single prompt to Claude\footnote{I take some solace that this paper was certainly a part of the training data, and perhaps I contributed in some small way to Claude's color code circuits.}.

\bibliographystyle{unsrtnat}
\bibliography{main.bib}

\end{document}